# Small Signal Capacitance in Ferroelectric HZO: Mechanisms and Physical Insights

Revanth Koduru[1], Atanu K. Saha[1], Martin M. Frank[2] and Sumeet K. Gupta[1]

*Abstract*— This study presents a theoretical investigation of the physical mechanisms governing small signal capacitance in ferroelectrics, focusing on Hafnium Zirconium Oxide ($Hf_{0.5}Zr_{0.5}O_2$, HZO). Utilizing a time-dependent Ginzburg Landau formalism-based 2D multi-grain phase-field simulation framework, we simulate the capacitance of metal-ferroelectric-insulator-metal (MFIM) capacitors. Our simulation methodology closely mirrors the experimental procedures for measuring ferroelectric small signal capacitance, and the outcomes replicate the characteristic butterfly capacitance-voltage behavior. We delve into the components of the ferroelectric capacitance associated with the dielectric response and polarization switching, discussing the primary physical mechanisms – domain bulk response and domain wall response – contributing to the butterfly characteristics. We explore their interplay and relative contributions to the capacitance and correlate them to the polarization domain characteristics. Additionally, we investigate the impact of increasing domain density with ferroelectric thickness scaling, demonstrating an enhancement in the polarization capacitance component (in addition to the dielectric component). We further analyze the relative contributions of the domain bulk and domain wall responses across different ferroelectric thicknesses. Lastly, we establish the relation of polarization capacitance components to the capacitive memory window (for memory applications) and reveal a non-monotonic dependence of the maximum memory window on HZO thickness.

This work was supported by NSF under grant number 2008412 and SRC under SRC LMD task 2959.001 and the Center for the Co-Design of Cognitive Systems (COCOSYS), one of seven centers in JUMP2.0, funded by SRC and DARPA. (*Corresponding author: Revanth Koduru*)

1) The authors are with the School of Electrical and Computer Engineering, Purdue University, West Lafayette, IN 47907 USA.

2) The authors are with IBM Research, Albany, NY 12203 USA.

## I. Introduction

Ferroelectric (FE) materials exhibit spontaneous polarization that can be switched by external electric fields exceeding their coercive field. Historically, this hysteretic property has been of great interest for non-volatile memory applications [1]. The recent discovery of ferroelectricity in doped-hafnium oxide ($HfO_2$) [2] has revitalized interest in FE devices. The CMOS compatibility of $HfO_2$ has led to the development of multiple flavors of FE devices for cutting-edge applications such as memory, computing in-memory (CiM), neuromorphic systems, and steep-slope transistors [3], [4], [5], [6], [7]. Further, the scale-free nature of ferroelectricity in $HfO_2$ [8], along with other appealing attributes [6], [9], [10], [11], has positioned FE devices as promising contenders for future electronics.

Another unique aspect of FE materials is their hysteretic and non-linear small-signal capacitive response to the applied voltage. This capacitive property, reflected in the butterfly capacitance-voltage (*C-V*) characteristics, has enabled various applications of FE materials, such as varactors, tunable filters, and oscillators [12], [13], [14], [15], [16]. Recently, researchers have leveraged the hysteretic FE capacitance for non-destructive sensing for CiM applications [17], [18], [19]. Given the broad range of applications of the hysteretic capacitive behavior of ferroelectrics, a deep understanding of the underlying physical mechanisms assumes great importance for proper application-driven device optimization.

Researchers have historically developed several models to capture and elucidate the FE *C-V* characteristics [16], [20-32]. One notable model is the modified Johnson's model [20], which extends the original Johnson's model [16] to the FE state. This empirical approach adapts the electric field dependence of the dielectric permittivity observed in paraelectric materials to ferroelectrics by shifting the origin of electric field dependence to the coercive field of the FE material. While effective in reproducing the experimental butterfly *C-V* characteristics, its empirical nature limits its ability to provide insights into the physical mechanisms governing the capacitance behavior.

Additionally, a class of approaches [23-27] proposed by various researchers relate the FE *C-V* characteristics to the displacement of underlying polarization domain walls (DWs). For instance, Jimenez et al. [25] extended Kittel's [26] approach of modeling DWs as rigid bodies moving under the action of an external electric field. In their work [25], the authors treated

DWs as stretched membranes under external electric fields and utilize their vibrational dynamics considering a linear restoration force. Employing this framework and incorporating the dependence of electric field on background permittivity, they derived equations for the butterfly *C-V* curves. However, these models are specifically tailored for small changes in the applied electric field and typically rely on the Preisach or other models to capture the large-signal hysteretic behavior. As a result, these models do not offer detailed insights into the relation between capacitive behavior, polarization switching and FE domain configurations.

Recent works by Massarotto and Segatto *et al.* [29-31] have addressed the long-standing gap between large-signal (LS) and small-signal (SS) capacitance characteristics of FE. Through experimental and simulation efforts, these studies have elucidated the differences between LS and SS behavior, and their relation to irreversible and reversible polarization switching [32]. In their simulation works [30], [31] based on Landau-Ginzburg-Devonshire theory, the authors present contrasting findings concerning the contribution of traps and FE response to the capacitance behavior. They further propose that domain wall motion might not significantly contribute to the FE capacitance response. However, it is noteworthy that these conclusions are contingent upon the assumptions regarding fixed domains and the absence of domain coupling utilized in the simulations.

Despite these advancements, a notable gap exists in understanding the physical mechanisms governing the FE small signal capacitance (SSC) characteristics. Specifically, the correlation of the capacitance to the FE polarization domain configurations and the phenomenon of capacitance increase well below the FE coercive voltage remains poorly understood. Further, a self-consistent framework capable of capturing both LS hysteresis and SS capacitance characteristics is yet to be developed. Such a framework would facilitate a more comprehensive understanding of the FE capacitance behavior and pave the way for extensive device optimizations.

To address these gaps, we present a multi-grain phase-field simulation framework based on time-dependent Ginzburg Landau (TDGL) formalism for metal-ferroelectric-insulator-metal (MFIM) capacitors. Our framework captures both the large-signal charge (*Q*)-voltage (*V*) hysteresis and the small-signal butterfly *C-V* characteristics. Our capacitance simulation methodology emulates the experimental measurement procedures for FE small signal capacitance (SSC) and unravels the physical mechanisms governing the *C-V* response. Focusing on Hafnium-Zirconium-Oxide ($Hf_{0.5}Zr_{0.5}O_2$) or HZO as the FE material, we explore different components of the capacitive response in the MFIM stacks.

Our analysis unveils two distinct responses behind the butterfly *C-V* characteristics:

1. Domain bulk response: the predominant response of regions within the bulk of FE domains.
2. Domain wall response: the response of regions in the domain walls of the FE layer, further comprising two sub-categories:
   - Wide domain wall response at the ferroelectric-

dielectric (FE-DE) interface.
   - Domain wall vicinity response.

We delve into the physical aspects of these mechanisms and their contributions to the capacitance characteristics, exploring their dependence on the applied bias voltage and the FE polarization domain configurations. Additionally, we investigate the impact of increasing domain density with scaling of ferroelectric thickness on these different components and their relative contribution to the total FE capacitance. We further analyze the effect of scaling FE thickness on the capacitive memory window for CiM applications.

The key contributions of this work include:

- Presenting a self-consistent framework capturing both the large signal and small signal FE characteristics
- Providing insights into the mechanisms governing the butterfly characteristics as well as the capacitance increase well below the coercive voltage of the FE.
- Correlating the FE capacitance to the underlying polarization switching mechanisms (domain growth and domain nucleation) and domain configurations.

## II. 2D MULTI-GRAIN PHASE-FIELD SIMULATION FRAMEWORK

We employ our in-house 2D multi-grain phase-field simulation framework (Fig. 1a), an extension of our previous works [33], [34], to simulate the FE small signal capacitance (SSC) characteristics in the MFIM stacks (Fig. 1c). This framework, based on the time-dependent Ginzburg Landau (TDGL) formalism and grain growth equation [36], captures the multi-domain *P*-switching in the FE layer while simultaneously incorporating the polycrystalline nature of HZO.

Given the significant computational cost of the phase-field simulations and the timescale of SSC measurements, we opt to perform the simulations in 2-dimensions (2D) rather than in 3-dimensions (3D). While transitioning to 2D may reduce the accuracy of the physical description of the MFIM capacitor compared to 3D simulations, we believe that using 2D simulations is reasonable. This expectation is based on our previous works and other studies that employ first principles calculations revealing an alternate polar-spacer layer (APSL) structure of HZO [8] in one direction along the cross-section and polar configuration in the other [35]. The lower gradient energy associated with APSL [35] stabilizes unit-cell wide domains and results in elastically independent *P*-switching along the APSL direction in HZO. Considering these factors, we simulate MFIM capacitors in 2D, accounting for the thickness and complete polar directions of HZO.

The polycrystalline structure of HZO is modeled using the grain growth equation (Fig. 1a.i) proposed by Krill et al. [36]. This equation (Eq. 1) describes the spatial and temporal evolution of the polycrystalline microstructure during crystallization, utilizing multiple abstract order parameters ($\eta_k, 0 \leq k \leq K$). The kinetics of these order parameters ($\eta_k$) is governed by Eq. 1, described below.

$$\frac{\partial \eta_k(r,t)}{\partial t} = -L\big(-a\eta_k(r,t) + b\eta_k^3(r,t)$$
$$+ 2c\eta_k(r,t)\sum_{s \neq k}^{K} \eta_s^2(r,t) - \kappa\nabla^2\eta_k(r,t)\big) \quad (1)$$

Here, $r$ represents spatial coordinates and $t$ represents time. The unitless parameters $a$=1, $b$=1, $c$=1, L=1, $\kappa$ = 0.5, and K=20 are calibrated to match the grain diameter distributions of simulated polycrystalline structures with experimental data [37] across different HZO film thicknesses. Detailed calibration results are presented in our earlier work [38].

The order parameters ($\eta_k$) represent variability among different grains in the polycrystalline structure. These grains can differ in various physical properties such as crystal orientation, material phase, stress and so on. In this study, we account for the inter-grain variability through spatial variations in the TDGL equation parameters, as discussed later.

For the 2D phase-field simulations, we utilize $x$-$z$ slices of the 3D polycrystalline structures generated by the grain growth equation (Fig. 1a. ii). These 3D structures have dimensions of 225 nm × 40 nm × 10 nm, and are sliced along the $y$-axis, resulting in 2D $x$-$z$ slices measuring 225 nm × 10 nm. Each 2D slice with unique grain configurations, serves as a distinct polycrystalline structure for the FE layer, representing one MFIM sample. This study encompasses analysis across 50 MFIM samples, each characterized by a different polycrystalline structure for the FE layer.

The 2D phase-field framework (Fig. 1a. iv) models the electrostatics and polarization ($P$) switching behavior in the MFIM stacks, accounting for the polycrystalline nature of the FE layer. The framework computes the potential $\phi$ and polarization $P$ profiles of the MFIM stack by solving the time-dependent Ginzburg Landau (TDGL) and Poisson's equations. These equations are iteratively and self-consistently solved in 2D real space using finite difference method with a grid spacing of 0.5 nm [38].

The time-dependent Ginzburg Landau (TDGL) equation governs the dynamics of polarization switching in the FE layer, relating the rate of change of polarization ($P$) to the total energy of the system ($F$) in its Euler-Lagrange form (Eq. 2) [39]

$$-\frac{1}{\Gamma}\frac{dP}{dt} = \frac{dF}{dP} \quad (2)$$

The total energy ($F$), includes free ($F_{free}$), electrostatic ($F_{elec}$) and $x$- and $z$-direction gradient ($F_{grad}$) energy components, with details of individual energy components in [39]. Substituting the individual energy components in Eq. 2 results in

$$-\frac{1}{\Gamma}\frac{\partial P}{\partial t} = \alpha_{gi}P + \beta_{gi}P^3 + \gamma_{gi}P^5 - g_{11}\frac{\partial^2 P}{\partial x^2} - g_{33}\frac{\partial^2 P}{\partial z^2} + \frac{d\phi}{dz} \quad (3)$$

Here, $\Gamma$ is the viscosity coefficient, $\alpha, \beta, \gamma$ are the Landau free energy parameters, and $g_{11}, g_{33}$ are the gradient energy coefficients along the $x$- and $z$- directions respectively.

Additionally, we account for the surface energy at FE-DE interface via the extrapolation length formalism [40], resulting in the boundary condition given by

$$\lambda \frac{\partial P}{\partial z} + P = 0 \quad (4)$$

where $\lambda$ is the extrapolation length.

We utilize the Landau parameters ($\alpha, \beta, \gamma$) to account for the inter-grain variability in the polycrystalline HZO layer. We introduce spatial variations in the values of Landau parameters between the individual grains. The values of $\alpha, \beta,$ and $\gamma$ for each grain (represented as $\alpha_{gi}, \beta_{gi},$ and $\gamma_{gi}$) are sampled from a Gaussian distribution around the mean values ($\alpha_0, \beta_0,$ and $\gamma_0$), with a standard deviation ($\sigma_0$), as shown in Fig. 1a.iii. The mean values and standard deviation are calibrated based on experimental $Q$-$V$ results.

Poisson's equation (Eq. 5) describes the electrostatic behavior of the MFIM system in terms of the electrostatic potential ($\phi$).

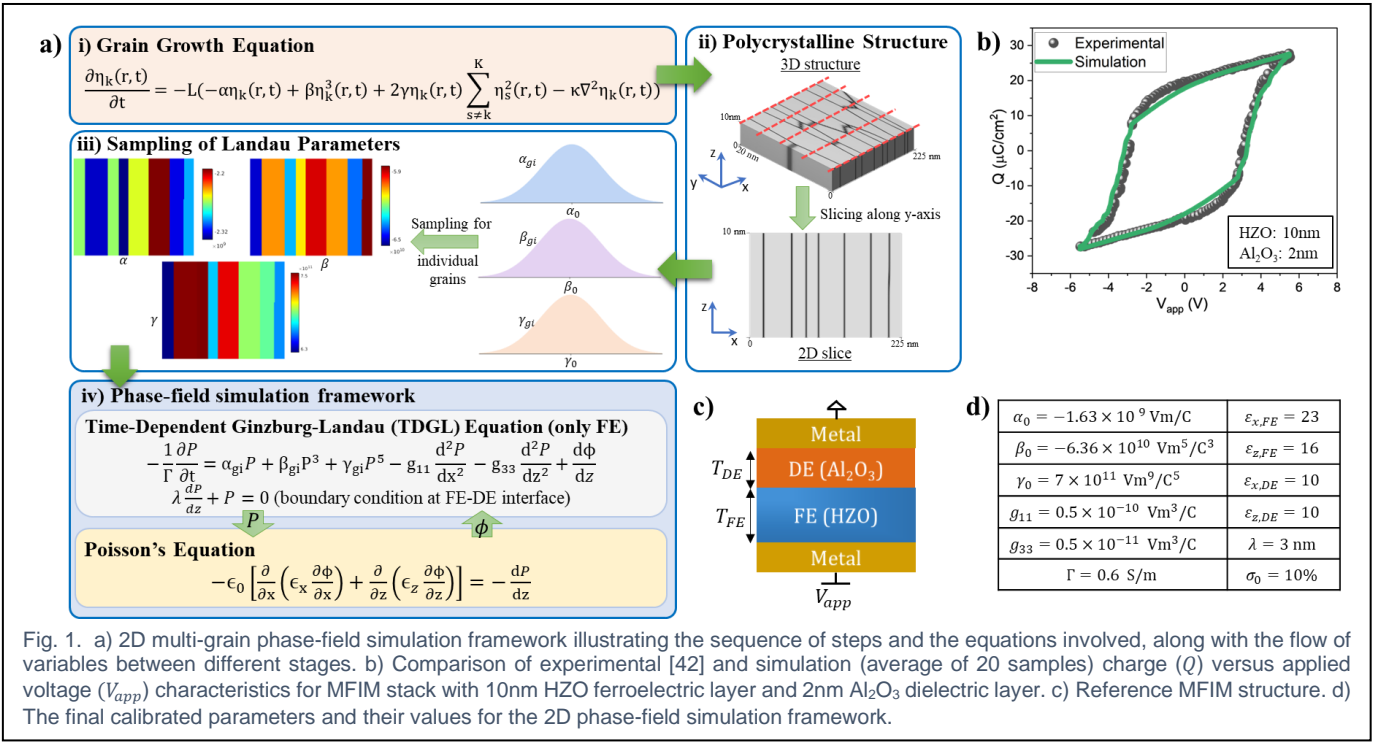

Fig. 1. a) 2D multi-grain phase-field simulation framework illustrating the sequence of steps and the equations involved, along with the flow of variables between different stages. b) Comparison of experimental [42] and simulation (average of 20 samples) charge ($Q$) versus applied voltage ($V_{app}$) characteristics for MFIM stack with 10nm HZO ferroelectric layer and 2nm Al$_2$O$_3$ dielectric layer. c) Reference MFIM structure. d) The final calibrated parameters and their values for the 2D phase-field simulation framework.

$$-\varepsilon_0\left[\frac{\partial}{\partial x}\left(\varepsilon_x\frac{\partial\phi}{\partial x}\right) + \frac{\partial}{\partial z}\left(\varepsilon_z\frac{\partial\phi}{\partial z}\right)\right] = -\frac{\partial P}{\partial z} \tag{5}$$

Here, $\varepsilon_0$ represents the vacuum permittivity, $\varepsilon_x$ and $\varepsilon_z$ represent the relative material permittivity in the $x$- and $z$- directions, respectively. The right-hand side of Eq. 5 accounts for the bound charges arising from the polarization gradient in the FE layer.

Our framework inherently captures the FE polarization switching via domain growth and domain nucleation mechanisms. It also incorporates polycrystalline structures with non-uniform grain shapes and sizes for the FE layer. However, to simplify this study, we consider solely the presence of orthorhombic phase of HZO and neglect leakage currents (reasonable for the thickness of MFIM stacks considered). We also assume uniform strain, no inter-grain elastic interactions [41], [42], [43] and trap/defect-free interfaces. As we will show later, the characteristic features of the small-signal butterfly $C$-$V$ curves can be obtained without invoking traps.

We calibrate our framework against experimental data, taken from Li et. al. [44], for the MFIM stacks consisting of 10 nm HZO FE layer and 2 nm Al$_2$O$_3$ DE layer. To ensure robust calibration and account for polycrystallinity-induced device-to-device variations, we calibrate by matching the average $Q$-$V$ characteristics of 20 simulated MFIM samples with experiments. It is important to note that, for the Landau parameters ($\alpha, \beta, \gamma$), calibration is performed on the mean value ($\alpha_0, \beta_0, \gamma_0$) and the standard deviation ($\sigma_0$) of the Gaussian distribution from which parameters of the individual grains are sampled.

For $\Gamma$, we utilize previously calibrated value of 0.6 S/m [33]. While this choice may affect the capacitance trends across high frequencies, we expect minimal impact on the capacitance and underlying physical mechanisms at the 1 MHz frequency considered in our study. The final calibrated parameters of the 2D multi-grain phase-field simulation framework are summarized in Fig. 1e, alongside a comparison of simulated and experimental $Q$-$V$ characteristics in Fig. 1b.

## III. FERROELECTRIC SMALL SIGNAL CAPACITANCE: SIMULATION METHODOLOGY

Utilizing the phase-field framework, we simulate the ferroelectric small signal capacitance (SSC) by closely replicating the experimental $C$-$V$ measurement methodologies [31]. We calculate the capacitance at any desired DC bias voltage ($V_0$) by simulating the MFIM stack under this bias $V_0$ until steady-state conditions are attained (Fig 2a). Subsequently, we superimpose a small-signal sinusoidal waveform of 1 MHz frequency and 1 mV amplitude onto the DC bias and simulate the MFIM stack under this combined waveform.

The charge response of the MFIM stack to the combined waveform depends on $V_0$ and consists of two components: reversible and irreversible responses [31], [32]. When the bias voltage falls in non-switching regions of the hysteretic $Q$-$V$ loop (e.g., $V_{01}$ in Fig 2b), far from the coercive voltage ($\pm V_C$) of the sample, the small-signal charge exhibits only a reversible response (Fig. 2d). This response is mainly due to the background permittivity of the materials and the oscillation of $P$-magnitude in the FE layer, without involving hysteretic $P$-switching.

However, when the bias voltage falls in the switching regions of the $Q$-$V$ loop (e.g., $V_{02}$ in Fig 2b), near $\pm V_C$ of the sample, the charge response involves both reversible and irreversible components. The response to the initial cycles of sinusoidal waveform shows an irreversible component due to hysteretic $P$-switching (inset for $V_{02}$ in Fig 2b). However, after a few cycles, the irreversible switching diminishes, leaving only the reversible component (Fig. 2c).

We calculate the small signal capacitance ($C$), considering solely the reversible charge response of the MFIM stack. The capacitance is calculated as the ratio of the reversible charge response ($Q_{sin}$) amplitude to the applied sinusoidal voltage ($V_{sin}$) amplitude, as given below.

$$C = \left.\frac{ampl(Q_{sin})}{ampl(V_{sin})}\right|_{V_0} \tag{6}$$

The butterfly $C$-$V$ curves (Fig. 3a) are then obtained by varying $V_0$ in steps of 50 mV, covering both the forward ($-5.5$ V to $5.5$ V) and backward ($5.5$ V to $-5.5$ V) paths, and calculating the capacitance at each step using Eq. 6 (Fig. 2a).

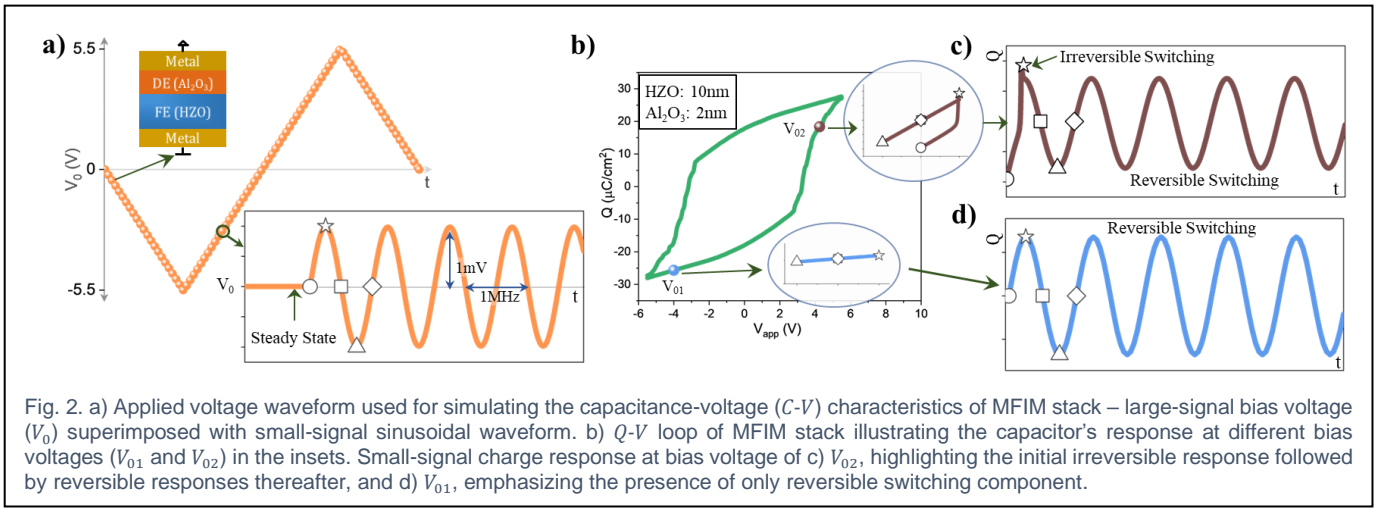

Fig. 2. a) Applied voltage waveform used for simulating the capacitance-voltage ($C$-$V$) characteristics of MFIM stack – large-signal bias voltage ($V_0$) superimposed with small-signal sinusoidal waveform. b) $Q$-$V$ loop of MFIM stack illustrating the capacitor's response at different bias voltages ($V_{01}$ and $V_{02}$) in the insets. Small-signal charge response at bias voltage of c) $V_{02}$, highlighting the initial irreversible response followed by reversible responses thereafter, and d) $V_{01}$, emphasizing the presence of only reversible switching component.

The small dimensions of the simulated MFIM samples (width of 225 nm) limit the number of grains in the FE layer. Consequently, we observe sharp step transitions in the $Q$-$V$ loops and multiple peaks and valleys in the $C$-$V$ curves. To capture smoother $C$-$V$ characteristics, we average $C$-$V$ responses across 50 MFIM samples, each characterized by a distinct polycrystalline HZO. This allows to account for the polycrystalline effects of HZO on the $C$-$V$ characteristics.

## IV. FERROELECTRIC SMALL SIGNAL CAPACITANCE: COMPONENTS

The simulated average $C$-$V$ results of the MFIM stack with 10 nm HZO and 2 nm $Al_2O_3$ (Fig. 3a) successfully replicate the butterfly characteristics observed in experiments [31], [44]. To elucidate the mechanisms governing these butterfly characteristics, we analyze the total capacitance by dividing it into dielectric and polarization capacitance components.

Let us approach this division from the perspective of total charge ($Q$) of the MFIM stack. It is evident that the HZO and $Al_2O_3$ layers, being in series, hold the same charge $Q$. In the HZO layer, $Q$ embodies the combined effect of two phenomena: the background permittivity response, referred to as the dielectric component ($Q_{de}$), and the response of the FE polarization domains to the electric field, referred to as the polarization component ($Q_P$). These components sum up to produce to total dielectric response in the $Al_2O_3$ layer. Note, we can attribute a portion of $Q$ in the $Al_2O_3$ layer as a response to the electric field/displacement of the dielectric component in the FE layer and the remaining portion as a response to the polarization component in the FE.

The dielectric ($Q_{de}$) and polarization ($Q_P$) components of $Q$ are calculated using the $z$-directed electric field ($E_{z,FE,avg}$) averaged along the $x$-direction in the FE layer [45].

$$Q_{de} = \varepsilon_0 \varepsilon_{z,FE} E_{z,FE,avg} \qquad (7)$$

$$Q_P = Q - Q_{de} \qquad (8)$$

$E_{z,FE,avg}$ is determined from the difference between the average potential at the top edge of the FE layer ($V_{FE,top,avg}$) i.e. the interface between HZO and $Al_2O_3$, and the bottom edge ($V_{FE,bot,avg}$) i.e. the interface between HZO and metal (Fig. 3c).

$$E_{z,FE,avg} = \frac{V_{FE,top,avg} - V_{FE,bot,avg}}{T_{FE}} \qquad (9)$$

Here, $T_{FE}$ is the thickness of the FE layer.

Analogous to total charge ($Q$), we divide the total capacitance ($C$) into two components: dielectric capacitance ($C_{de}$) and polarization capacitance ($C_P$). These components are calculated using the response of corresponding charge components to the sinusoidal waveform ($Q_{de,sin}, Q_{P,sin}$).

$$C_{de} = \frac{ampl(Q_{de,sin})}{ampl(V_{sin})}\bigg|_{V_0} \qquad (10)$$

$$C_P = \frac{ampl(Q_{P,sin})}{ampl(V_{sin})}\bigg|_{V_0} \qquad (11)$$

$C_{de}$ exhibits an inverted butterfly shape (Fig. 3b), while $C_P$ displays the characteristic butterfly shape (Fig. 3b) and is mainly responsible for the overall butterfly characteristics.

In the following subsections, we discuss the details of these capacitance components and highlight the underlying physical mechanisms. For this, we utilize the spatial profiles of polarization $P(x,z)$ and electric-field $E(x,z)$ of a representative MFIM sample at different bias voltages. These profiles (shown in subsequent figures) focus on width of 100 nm to provide a clear depiction of the $P$-domains and $E$-field lines.

## V. THE DIELECTRIC CAPACITANCE COMPONENT

The dielectric capacitance component ($C_{de}$) arises from the background permittivity response of the MFIM stack to the applied sinusoidal waveform. This component (from Eq. 7 and 10) depends on the out-of-plane or $z$-directed electric field in MFIM and exhibits an inverted butterfly shape (Fig. 4a).

To understand this inverted butterfly behavior, let us briefly discuss the formation of multi-domains in the FE layer (Fig. 3c) and its impact on $E$-field distribution. Imperfect screening of the polarization charges at the FE-DE interface generates an electric field opposite to the $P$ direction in the FE layer, known as the depolarization field. This depolarization field increases the electrostatic energy of the system. Ferroelectrics often minimize this energy increase by breaking into multiple domains with opposite $P$ directions (Fig. 3c), albeit at the cost of increased gradient energy [39]. These multi-domains compensate the $E$-field lines originating from one domain in the adjacent oppositely polarized domains. This leads to in-plane $E$-fields or stray fields near domain walls (DW) at the FE-DE interface and reduces the out-of-plane $E$-field (Fig. 3c) and electrostatic energy [39].

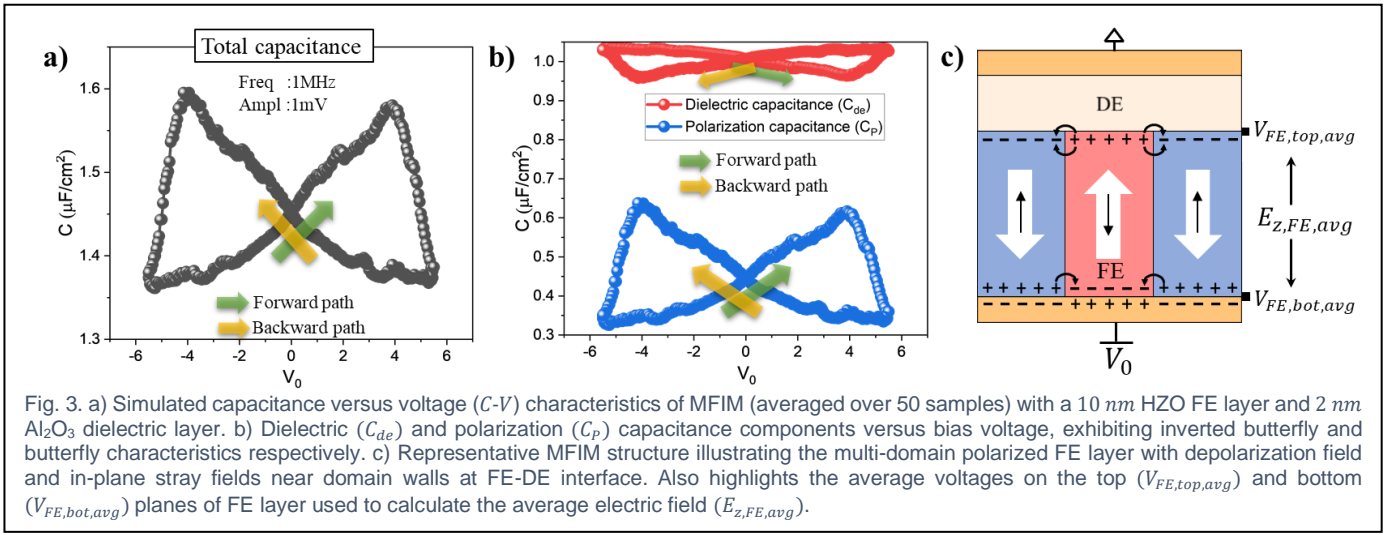

Fig. 3. a) Simulated capacitance versus voltage ($C$-$V$) characteristics of MFIM (averaged over 50 samples) with a 10 nm HZO FE layer and 2 nm $Al_2O_3$ dielectric layer. b) Dielectric ($C_{de}$) and polarization ($C_P$) capacitance components versus bias voltage, exhibiting inverted butterfly and butterfly characteristics respectively. c) Representative MFIM structure illustrating the multi-domain polarized FE layer with depolarization field and in-plane stray fields near domain walls at FE-DE interface. Also highlights the average voltages on the top ($V_{FE,top,avg}$) and bottom ($V_{FE,bot,avg}$) planes of FE layer used to calculate the average electric field ($E_{z,FE,avg}$).

The in-plane $E$-field component in the MFIM is proportional to the number of DWs in the FE layer. Moreover, its strength is highest when the magnitudes of $+P$ and $-P$ domains across DW is equal [46]. As the magnitudes of these domains vary with changing bias voltage, we observe transformation of some in-plane $E$-fields to out-of-plane direction and vice-versa at different bias voltages, depending on the relative $P$-magnitudes in neighboring domains. Note that, $C_{de}$ is governed by the out-of-plane $E$-fields and their induced charges on the metal electrodes. Thus, larger in-plane $E$-fields result in smaller $C_{de}$.

Now, let us discuss $C_{de}$ versus $V_0$ focusing on the forward voltage path, where $V_0$ increases from $-5.5$ V to 5.5 V (Fig. 4a). Starting at $-5.5$ V, the high negative voltage stabilizes the majority of FE layer in $-P$ (blue regions in Fig. 4b), with a smaller portion exhibiting $+P$ (red regions in Fig. 4b). This leads to fewer domains and DWs in the FE layer. The negative bias voltage also induces an asymmetry in the domains, with the magnitude of $-P$ greater than that of the $+P$ domains. This asymmetry, coupled with the fewer DWs, results in a predominantly out-of-plane electric field in the MFIM stack, with minimal in-plane components (inset of Fig. 4b). Consequently, we observe a large $C_{de}$ of the MFIM stack.

As $V_0$ gradually increases, the magnitude of $+P$ increases while the magnitude of $-P$ domains decrease. This reduction in asymmetry between $+P$ and $-P$ domains (Fig. 4c) allow for more $E$-field compensation near the DWs. In other words, some out-of-plane $E$-fields transform to in-plane $E$-fields (op-to-ip transformation), leading to an increase in the in-plane $E$-field component (inset in Fig. 4c). Consequently, we observe a decrease in $C_{de}$ with increasing $V_0$ (Fig. 4a).

As $V_0$ increases further and approaches coercive voltage ($+V_C$) of the sample, FE layer undergoes $P$-switching through domain growth and domain nucleation (Fig. 4e). The nucleation of new $+P$ domains increases the number of domains and DWs in the FE layer. This enhances the in-plane $E$-field at FE-DE interface, leading to a continued decrease in $C_{de}$ (Fig. 4a).

As $V_0$ increases beyond $+V_C$, the $-P$ domains become thinner and collapse, causing the coalescence of $+P$ domains (Fig. 4d). This reduces the number of DWs in the FE layer and decreases the in-plane $E$-field while increasing the out-of-plane $E$-field (ip-to-op transformation). This change in the $E$-field distribution in the MFIM increases $C_{de}$ for $V_0 > +V_C$ (Fig. 4a).

Similar mechanisms, but in the opposite direction, govern $C_{de}$ along the backward voltage path. As $V_0$ decreases from 5.5 V, the reduction in asymmetry between $+P$ and $-P$ domains increases the in-plane $E$-fields and decrease $C_{de}$. Subsequently, as $V_0$ decreases further, aided by the nucleation of $-P$ domains around $-V_C$, the in-plane $E$-field continues to increase, further reducing $C_{de}$. However, as $V_0$ decreases below $-V_C$, $-P$ domains coalesce, reducing in-plane $E$-field and increasing the out-of-plane $E$-field, which in turn increases $C_{de}$. This interplay between in-plane and out-of-plane $E$-field transformations, $P$-switching via domain nucleation and coalescence, and the underlying $P$-domain configurations results in the inverted butterfly characteristics of the dielectric capacitance component shown in Fig. 4a.

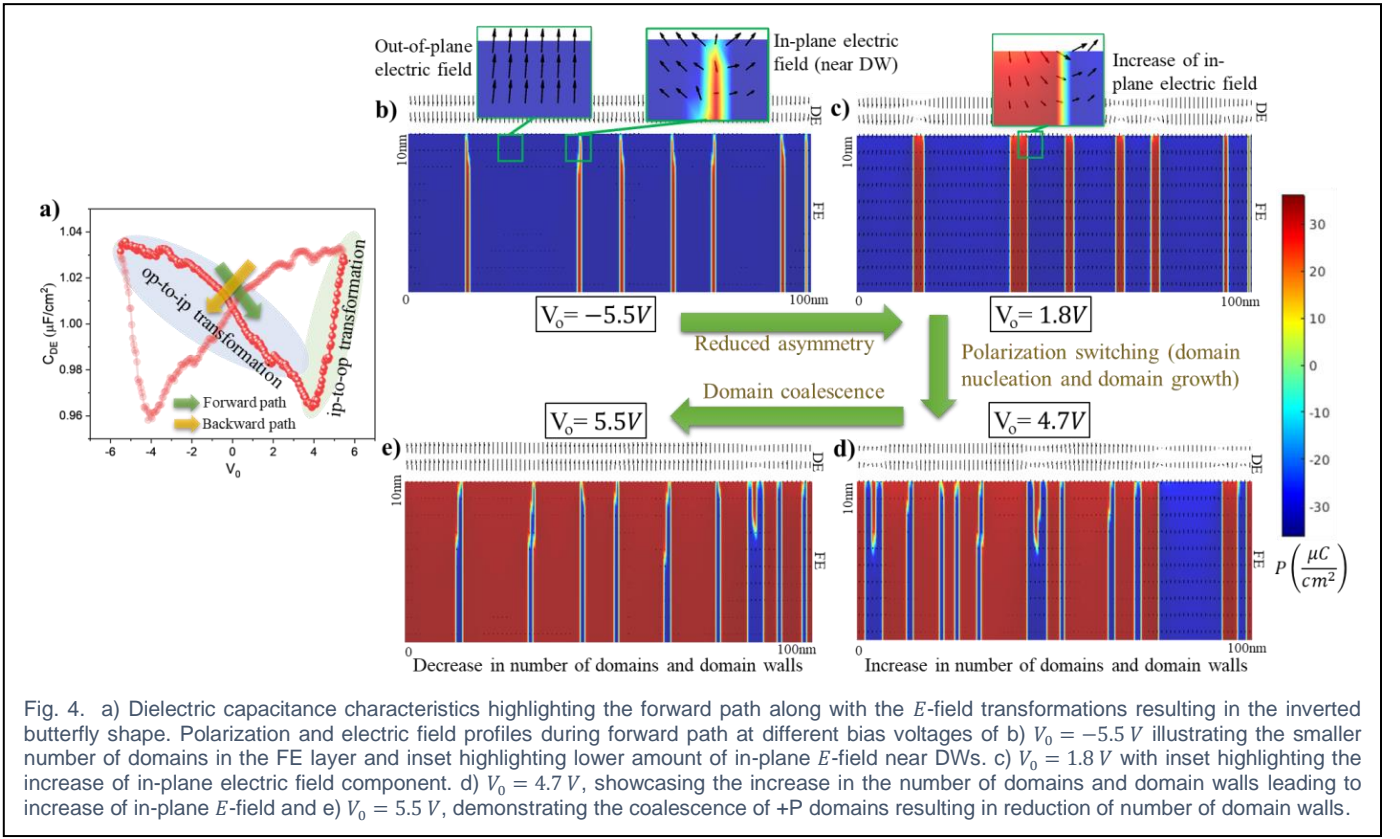

Fig. 4. a) Dielectric capacitance characteristics highlighting the forward path along with the $E$-field transformations resulting in the inverted butterfly shape. Polarization and electric field profiles during forward path at different bias voltages of b) $V_0 = -5.5\ V$ illustrating the smaller number of domains in the FE layer and inset highlighting lower amount of in-plane $E$-field near DWs. c) $V_0 = 1.8\ V$ with inset highlighting the increase of in-plane electric field component. d) $V_0 = 4.7\ V$, showcasing the increase in the number of domains and domain walls leading to increase of in-plane $E$-field and e) $V_0 = 5.5\ V$, demonstrating the coalescence of +P domains resulting in reduction of number of domain walls.

## VI. The Polarization Capacitance Component

The polarization capacitance ($C_P$) arising from the response of FE polarization to the sinusoidal waveform exhibits a butterfly shape (Fig. 3b). Our analysis reveals two physical responses governing $C_P$ characteristics across $V_0$:

1. Domain bulk response: This refers to the response of the FE regions deep within the polarization domains, away from the domain walls (DWs).
2. Domain wall response, which comprises:
   a. Wide domain wall response: The response of wide domain walls or the "softer" domains near the FE-DE interface.
   b. Domain wall vicinity response: This refers to the response of the FE regions adjacent to DWs.

Given the small amplitude of the sinusoidal voltage (1mV), it is difficult to discern the small signal response in the polarization profiles of the FE layer. To effectively illustrate the small changes in $P$, we utilize the polarization-amplitude ($ampl$-$P$) profiles (Fig. 5b) that depict the minute polarization response to the small-signal voltage. For any bias voltage, the polarization-amplitude profiles are obtained by spatially subtracting the polarization at the time instance corresponding to zero sinusoidal voltage ($t_0$) from the polarization at the time instance corresponding to the peak of the sinusoidal voltage ($t_{max}$), as in Fig. 5a and Eq. 12.

$$ampl\text{-}P(x,z) = P(x,z,t_{max}) - P(x,z,t_0) \qquad (12)$$

To quantify the contributions of the above-mentioned responses, we define averaged domain bulk capacitance ($C_{DB}$) and averaged domain wall capacitance ($C_{DW}$), which capture the domain bulk and domain wall responses, respectively. Given the spatially distributed nature of these responses, we partition FE layer into domain bulk (DB) and domain wall (DW) regions. DW regions are characterized by lower $P$-magnitude and higher gradient energy compared to the DB regions. In the considered MFIM configuration, we typically observe hard or sharp DWs [39] with no lattice points in the $P$ transition region. However, for calculating $C_{DW}$, we include the lattice points on either side of transition into the DW region, resulting in a DW width of 2 lattice points for hard DWs. Under certain conditions (discussed in Section VI.B.1), softer DWs with a greater width than 2 lattice points are observed.

The averaged capacitances ($C_{DB}$ and $C_{DW}$) are calculated at each bias voltage by spatially aggregating the polarization-amplitude ($ampl$-$P$) of the DB and DW regions. This aggregated value is then normalized with the amplitude of the sinusoidal waveform ($ampl(V_{sin})$) and the area of FE layer in $x$-$z$ plane ($Area$), as follows:

$$C_{DB} = \frac{\sum_z \sum_x ampl\text{-}P_{DB}(x,z)}{ampl(V_{sin})} * \frac{1}{Area} \qquad (13)$$

$$C_{DW} = \frac{\sum_z \sum_x ampl\text{-}P_{DW}(x,z)}{ampl(V_{sin})} * \frac{1}{Area} \qquad (14)$$

Here, $ampl\text{-}P_{DB}(x,z)$ and $ampl\text{-}P_{DW}(x,z)$ represent the polarization-amplitude of the domain bulk and domain wall regions, respectively. $ampl(V_{sin}) = 1$ mV and $Area = 225$ nm $\times 10$ nm.

### A. Domain bulk (DB) response

Domain bulk regions typically exhibit low gradient energy and comprises either $+P$ or $-P$ domains. The minimal gradient energy implies that the polarization of DB regions is governed by the Landau-Khalatnikov (LK) equation associated with $F_{free}$ and $F_{elec}$ of the FE material (Eq. 15)

$$E = \alpha P + \beta P^3 + \gamma P^5 \qquad (15)$$

The LK equation exhibits S-shaped curve with segments of varying slopes: low-slope, moderate-slope, and high-slope segments (marked in Fig. 6c). The response of the DB regions to the small-signal waveform depends on the slope of the LK curve segment they fall into.

To illustrate this dependence, we select certain reference lattice points across the DB regions in the FE layer (green stars in Fig. 6a). We examine the polarization ($P$) of these reference points versus the $E$-field they experience, obtained from the phase-field model. We refer to these as the $P$-$E$ positions of these points and plot them against the S-shaped LK curve (Fig. 6c). Note that the LK curve is constructed using the mean valued Landau parameters ($\alpha_0, \beta_0, \gamma_0$), while the actual $P$-$E$ positions account for the polycrystallinity-induced variability in the Landau parameters. The slope of LK curve corresponding to these $P$-$E$ positions determine the response of DB regions to the small-signal waveform.

Let us begin by focusing on the forward voltage path starting at $V_0 = -5.5$ V. At this $V_0$, FE layer consists mostly of $-P$

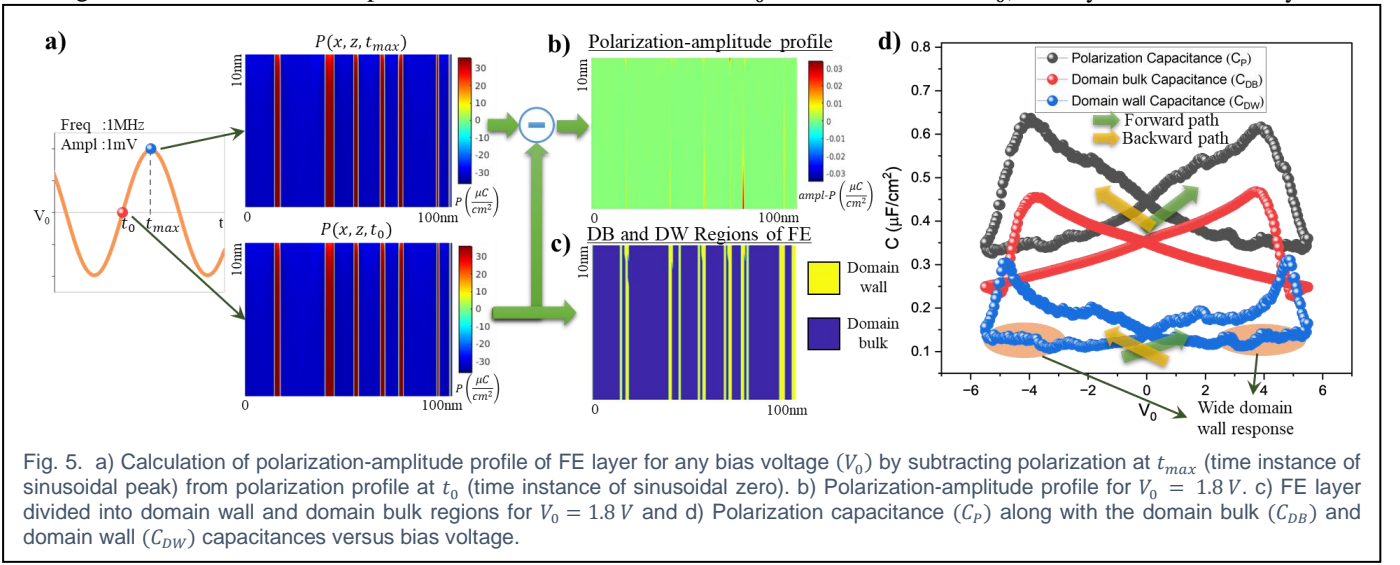

Fig. 5. a) Calculation of polarization-amplitude profile of FE layer for any bias voltage ($V_0$) by subtracting polarization at $t_{max}$ (time instance of sinusoidal peak) from polarization profile at $t_0$ (time instance of sinusoidal zero). b) Polarization-amplitude profile for $V_0 = 1.8\,V$. c) FE layer divided into domain wall and domain bulk regions for $V_0 = 1.8\,V$ and d) Polarization capacitance ($C_P$) along with the domain bulk ($C_{DB}$) and domain wall ($C_{DW}$) capacitances versus bias voltage.

regions with thin $+P$ domains (Fig. 6a). The $P$-$E$ positions of the reference DB points (Fig. 6c $-5.5$ V) fall into the low-slope segments of the LK curve. This indicates minimal response of these DB regions to the sinusoidal waveform, which is evident from the low $ampl$-$P$ response, approximately $10^{-4}$ μC/cm$^2$, as shown in Fig. 6b. However, due to the DB regions extending across almost the entire FE, their collective contributes significantly to the polarization capacitance (Fig. 5d).

As $V_0$ increases from $-5.5$ V but remains below $+V_C$, the $P$-switching is dominated by domain growth (Fig. 6d for 1.8 V). The increase in $V_0$ increases the polarization of $+P$ domains, shifting their $P$-$E$ positions further into the low-slope segment of the LK curve (Fig. 6c). Consequently, the response of $+P$ DB regions to the sinusoidal waveform decreases. On the other hand, the magnitude of $-P$ domains decreases, shifting their $P$-$E$ positions towards the high-slope segment or the turnaround point of the LK curve (Fig. 6c). This increases their response to the sinusoidal waveform to around $10^{-3}$ μC/cm$^2$ (Fig. 6e). Overall, $C_{DB}$ increases with $V_0$ (Fig. 5d) as the majority of FE is negatively polarized.

The increase in $C_{DB}$ continues with further increase in $V_0$ as the $P$-$E$ positions of $-P$ domains move further towards the high-slope segment of the LK curve. As $V_0$ approaches the vicinity of $+V_C$, the $P$-$E$ positions of $-P$ domains reach the turnaround point of the LK curve. As a result, even a slight increase in $V_0$ causes these $-P$ regions to switch to $+P$. However, the switching mechanism depends on the proximity of these $-P$ regions to DWs. Regions near DWs switch via domain growth and the capacitive response of these regions is discussed in Section VI.B.2 (on domain wall vicinity response). The regions away from DWs switch via domain nucleation, where new $+P$ domains nucleate from the FE-DE interface.

The nucleation of $+P$ domains relocates the $P$-$E$ positions of these regions to the low-slope segment of the LK curve (Fig. 6h), leading to a drop in their response to the sinusoidal waveform. Additionally, nucleation increases the number of domains and DWs in the FE layer, reducing the domain bulk (Fig. 6f). Thus, after nucleation, these two factors lead to a drop in $C_{DB}$, which was previously increasing with $V_0$. However, due to averaging over 50 MFIM samples with polycrystalline induced variations in $V_C$, we observe a relatively smooth decrease in $C_{DB}$ with $V_0$.

With further increase in $V_0$, the $+P$ domains grow and coalesce together, reducing the area of DWs and increasing the DB regions in the FE layer (Fig. 6g). However, the $P$-$E$ positions of DB regions now traverse along low-slope segment of the LK curve (Fig. 6i). Due to these combined effects, we observe only a slight increase in $C_{DB}$ at higher bias voltages (around 5 V in Fig. 5d).

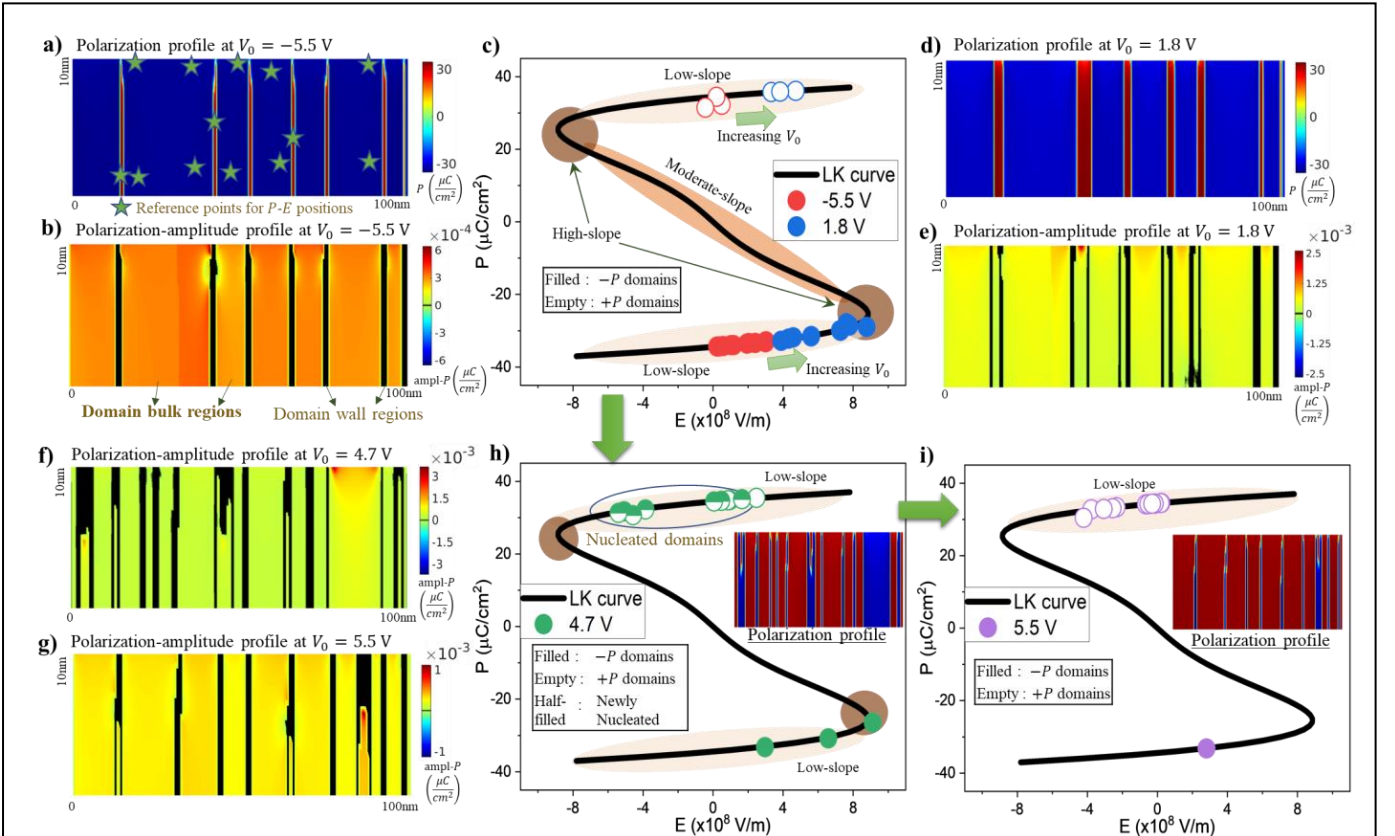

Fig. 6. a) Polarization profile $P(x,z)$ at $V_0 = -5.5$ V, highlighting the reference points whose $P$-$E$ positions are analyzed. b) Polarization-amplitude profile $ampl$-$P(x,z)$ at $V_0 = -5.5$ V emphasizing lower response of domain bulk regions; black regions represent domain walls. c) $P$-$E$ positions of DB reference points plotted against LK curve for $V_0 = -5.5$ V and 1.8 V, showing traversal of $+P$ domains along low-slope and $-P$ domains towards high-slope segments of LK curve. d) Polarization and e) polarization-amplitude profiles at $V_0 = 1.8$ V, highlighting increased response of the DB regions to sinusoidal waveform. $ampl$-$P(x,z)$ at f) $V_0 = 4.7$ V showing reduced DB area due to nucleation, and g) $V_0 = 5.5$ V showing increased DB area due to domain coalescence. h) $P$-$E$ positions of DB reference points for $V_0 = 4.7$ V, indicating the shift from high-slope to low-slope segment of LK curve due to domain nucleation, $P(x,z)$ in inset. i) $P$-$E$ positions of DB reference points for $V_0 = 5.5$ V, indicating the shift along low-slope segment of LK curve with $P(x,z)$ in inset.

Similar mechanisms involving the opposite polarization govern the domain bulk capacitance in the backward path. As $V_0$ decreases from +5.5 V, the $+P$ DB regions move towards the high-slope segment of the LK curve, increasing $C_{DB}$. Subsequently, the switching of $+P$ domains to $-P$ domains via domain nucleation reduces the area of DB regions and $C_{DB}$. This is followed by a slight increase in $C_{DB}$ due to the coalescence of $-P$ domains. These interactions, such as the traversal of the LK curve by the DB regions, polarization switching via domain nucleation, and domain coalescence, dictate the domain bulk capacitance ($C_{DB}$) and its dependency on bias voltage (Fig. 5d).

## B. Domain wall response

Domain wall refers to the transition region between different polarization directions, in this case, $+P$ and $-P$ domains. These DW regions are characterized by high gradient energies due to spatial variations in polarization and a lower (closer to zero) $P$-magnitude than the DB regions. Our analysis reveals two different DW responses to the sinusoidal waveform: wide DW response at the FE-DE interface and DW vicinity response.

As mentioned earlier, for the simulated MFIM structure, we typically observe hard DWs with a considered width of 2 lattice points. The response of these regions to the sinusoidal waveform is discussed in Section VI.B.2 on domain wall vicinity response. However, for certain $V_0$, we observe the formation of wide or "softer" DWs near the FE-DE interface. The width of these DWs is greater than 2 lattice points and we refer to the response of these regions as a wide domain wall response at the FE-DE interface, which we discuss next.

### 1) Wide domain wall response at the FE-DE interface

Domain walls typically widen at the FE-DE interface than in the bulk of the FE layer [47]. However, at high negative (positive) bias voltages where $+P$ ($-P$) domains become very thin, we observe further widening of the DWs at the FE-DE interface. These widened DWs, due to their distinct properties, exhibit heightened response to the sinusoidal voltage, referred to as the wide domain wall response at the FE-DE interface (highlighted in Fig. 5d).

First, let us discuss the typical DW widening at the FE-DE interface. Wide DWs at the interface imply a slower transition of polarization to the domain bulk (DB) values compared to the FE bulk (away from the interface), as shown in Fig. 7c. In the FE bulk, DWs are sharp (hard), with $P$-magnitude on either side of DW close to DB values. However, at the interface, stray or in-plane $E$-fields near the DWs (discussed in section. V, Fig. 3c) reduce the out-of-plane $E$-field, resulting in lower polarization magnitudes. These regions create an additional transition zone between the $+P$ and $-P$ domains, widening the DWs. This phenomenon is observable at any bias voltage ($V_0$) and has been investigated in prior works [47].

At high negative (positive) bias voltages, this DW widening at the interface is heightened due to thin $+P$ ($-P$) domains and strong depolarization fields. For instance, at $V_0 = -5.5$ V, thinning of $+P$ domains (Fig. 7d) reduces the $P$-magnitude in the $+P$ domain bulk [46], even away from the FE-DE interface. At the interface, this reduction of $+P$ magnitude is intensified due to the polarization gradient along $z$-direction (Fig. 7b) induced by the depolarization fields. The lower $+P$ domain magnitude at the interface slows down the transition between the $+P$ and $-P$ domains, leading to further widening of the DWs. As a result, we observe conical $+P(-P)$ domains (Fig. 7d) at high negative (positive) voltages, with DWs near the interface exhibiting slightly "softer" characteristics than those in the FE bulk.

With this understanding, let us explore the response of these wide DWs to the sinusoidal voltage. Focusing on the forward voltage path, we examine the $P$-$E$ positions of two reference lattice points (green stars in Fig. 7d): one in the wide DW region and the other in the $+P$ domain near the interface. At $V_0 = -5.5$ V, the polarization of the wide DW region falls in the transition between the $+P$ and $-P$ values, with its $P$-$E$ position in the moderate-slope segment of the LK curve (circle in Fig. 7f). In contrast, the $P$-$E$ positions of $+P$ domains at the interface fall along the high-slope segment (square in Fig. 7f), owing to their slightly lower $P$-magnitude than the FE bulk (away from the interface). Note that, the $P$-$E$ positions deviate from the LK curve due to the high gradient energy associated with the DW and FE-DE interface regions.

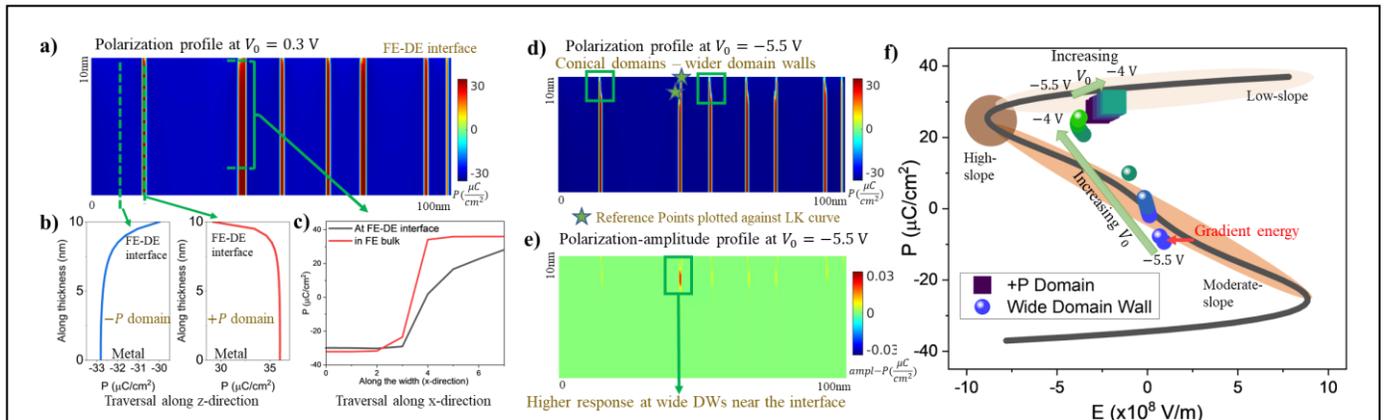

Fig. 7. a) Polarization profile at $V_0 = 0.3$ V with b) line plot of polarization along the thickness ($z$-direction) in $+P$ and $-P$ domains illustrating the reduced polarization magnitude at the FE-DE interface and the polarization gradient along the $z$-direction, c) line plot of polarization along the $x$-direction crossing a domain wall showing the widening of DW at the FE-DE interface compared to the FE bulk. d) Polarization profile at $V_0 = -5.5$ V highlighting the formation of conical domains at the interface and the reference points of plotted $P$-$E$ positions in green stars. e) Polarization- amplitude profile for $V_0 = -5.5$ V with higher response at the wide DWs near the interface. f) Reference points of $+P$ domain (square) and wide DW region (circle) plotted against the LK curve for $V_0$ increasing from $-5.5$ V to $-4$ V, showing the traversal of wide DW point along the moderate-slope to high-slope segment and the $+P$ domain point away from the high-slope segment.

The $P$-$E$ positions of the wide DW and $+P$ domains along the moderate and high-slope segments, respectively, imply that these regions exhibit heightened response to the sinusoidal waveform. This heightened response, as evident from the $ampl$-$P$ profile (Fig. 7e), is approximately $10^{-2}$ $\mu C/cm^2$, compared to DB regions with responses around $10^{-4}$ $\mu C/cm^2$. However, since these regions constitute only a small portion of the FE layer, their aggregate response ($C_{DW}$) is smaller than the aggregated domain bulk response ($C_{DB}$), as observed in Fig. 5d.

As $V_0$ increases from $-5.5$ V, the polarization of $+P$ domains increase, shifting their $P$-$E$ positions towards the low-slope segment of the LK curve (squares in Fig. 7f). Simultaneously, the $P$-$E$ positions of wide DW regions at the FE-DE interface traverse along the moderate slope segment towards the high-slope turnaround point (circles in Fig. 7f). However, due to the very small area of these wide domain walls, domain wall capacitance ($C_{DW}$) shows minimal variation with increasing the bias voltage ($V_0$).

With the continued increase in $V_0$, the $+P$ domains stabilize, reducing the domain wall widening at the FE-DE interface. This stabilization transforms the conical $+P$ domains into a more cylindrical shape. As a result, the previously softer (wide) DW regions at the FE-DE interface convert to $+P$ domains and transition to hard DWs. Hence, their $P$-$E$ positions reach the low-slope segment of the LK curve (circle in Fig. 7f for $V_0 = -4V$), reducing their response to the sinusoidal waveform. This causes a slight dip in the domain wall capacitance ($C_{DW}$), as highlighted in Fig. 5d. With further increase in $V_0$, the regions in the vicinity of the domain walls begin to respond to the sinusoidal waveform, which will be discussed in the next section.

Similarly, at high positive bias voltages ($V_0 = 5.5$ V), we observe the formation of conical $-P$ domains with wider domain walls at the FE-DE interface. These wider DWs and $-P$ domains near the interface regions exhibit higher response to the sinusoidal waveform. However, as $V_0$ decreases, these conical $-P$ domains transform into cylindrical domains, leading to a dip in the domain wall capacitance ($C_{DW}$). In summary, the formation of softer domains at the FE-DE interface with wider domain walls governs the wide-domain wall response at highly positive and negative bias voltages.

### 2) Domain wall vicinity response

We define the domain wall vicinity response as the response of the FE regions adjacent to the $P$-transition regions, specifically the 2 lattice points considered in the DW. This response is present across all bias voltages ($V_0$) whenever DWs are present in the FE layer and is a major contributor to the domain wall ($C_{DW}$) and the polarization ($C_P$) capacitances. The magnitude of this contribution depends on the area of DWs in the FE layer, and is significantly influenced by the underlying domain configurations, $P$-switching mechanisms, and bias voltage.

To understand these dependencies, let us focus on the forward voltage path and the region surrounding a representative DW in the FE layer (dashed green line in Fig. 8a). Due to the hard DW, the polarization in this region shows the sharp spatial transition from $+P$ to $-P$ domains (black line in Fig. 8b). Let us consider the lattice points along this region (markers in Fig. 8b) and examine their $P$-$E$ positions. We will mainly focus on the lattice points on either side of the transition region (green and purple points in Fig. 8d), concentrate on the range of $V_0$ (2.3 to 2.35 V for the sample being discussed) over which this DW undergoes motion or domain growth. At other bias voltages, the *P-E* positions of these points typically fall into the low-slope segment of the LK curve.

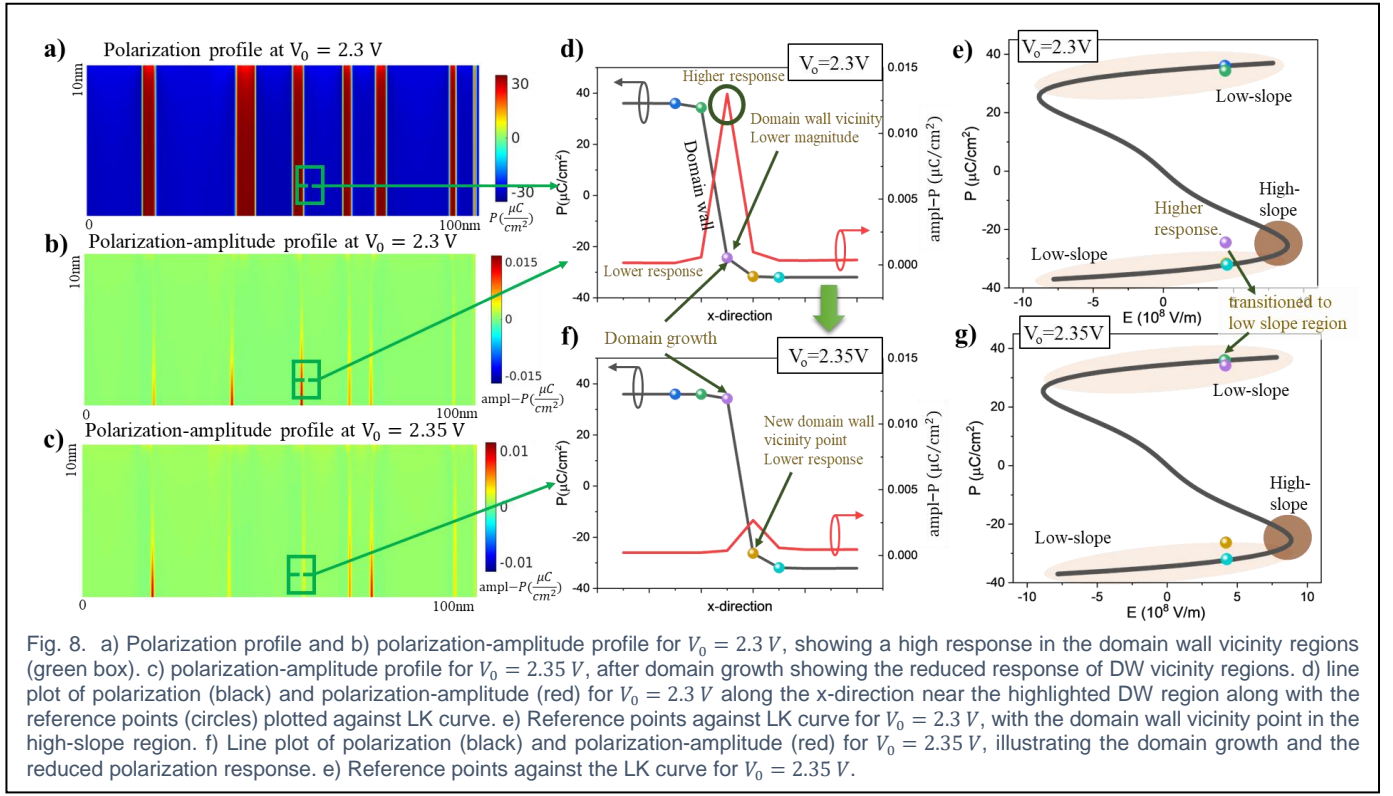

Fig. 8. a) Polarization profile and b) polarization-amplitude profile for $V_0 = 2.3$ $V$, showing a high response in the domain wall vicinity regions (green box). c) polarization-amplitude profile for $V_0 = 2.35$ $V$, after domain growth showing the reduced response of DW vicinity regions. d) line plot of polarization (black) and polarization-amplitude (red) for $V_0 = 2.3$ $V$ along the x-direction near the highlighted DW region along with the reference points (circles) plotted against LK curve. e) Reference points against LK curve for $V_0 = 2.3$ $V$, with the domain wall vicinity point in the high-slope region. f) Line plot of polarization (black) and polarization-amplitude (red) for $V_0 = 2.35$ $V$, illustrating the domain growth and the reduced polarization response. e) Reference points against the LK curve for $V_0 = 2.35$ $V$.

At $V_0 = 2.3$ V, the *P-E* position of $+P$ lattice point (green marker) is along the low-slope segment of the LK curve (Fig. 8e), resulting in minimal response to the sinusoidal waveform (Fig. 8d). In contrast, the *P-E* position of $-P$ lattice point (purple marker) is along the high-slope segment (Fig 8e). Consequently, this $-P$ lattice point exhibits heightened response, of around $1.5 \times 10^{-2}$ $\mu C/cm^2$, to the sinusoidal waveform (Fig. 8b, red line in Fig. 8d). This asymmetry in the polarization response on the two sides of the DW leads to reversible change in the DW position in response to the sinusoidal voltage and is often referred to as domain wall vibration in literature [23], [24], [25], [27].

With a slight increase in $V_0$ to 2.35 V, the purple marker (exhibiting heightened response at 2.3V) transitions to $+P$ via domain growth or the domain wall motion (Fig. 8f). This transition shifts its *P-E* position from the high-slope segment to the low-slope segment (purple marker in Fig. 8g). Consequently, this lattice point shows reduced response to sinusoidal waveform (purple marker in Fig. 8f). Additionally, the new $-P$ lattice point (gold marker in Fig. 8f) adjacent to the transition region is still slightly away from the high-slope turnaround point (Fig. 8g). Due to the significant slope variation of the LK curve near the turnaround point, this gold lattice point exhibits a slightly lower sinusoidal response (Fig. 8f), around $2 \times 10^{-3} \mu C/cm^2$. Consequently, the small signal charge response in the vicinity of this DW diminishes at 2.35 V (Fig. 8c, f).

However, there exist numerous DWs across the 50 different MFIM samples and the polycrystalline variations introduce variability in the voltages at which these different DWs respond. Due to the consideration of Gaussian distribution of Landau parameters, we observe a Gaussian distribution for the voltages at which different DWs undergo DW motion. Consequently, as $V_0$ increases, more DWs approach the verge of undergoing the DW motion leading to a continuous increase in $C_{DW}$ with $V_0$ (Fig. 5d).

As $V_0$ approaches the vicinity of the coercive voltage ($+V_C$), FE layer undergoes *P*-switching via domain nucleation (Fig. 4d). This process increases the number of $+P$ domains and DWs in the FE layer (Fig. 6f), increasing the domain wall vicinity response. This is evident from the rise of $C_{DW}$ *after* the peak of domain bulk capacitance (which occurs before nucleation) in Fig. 5d. However, as $V_0$ increases further, $+P$ domains grow and coalesce with each other (Fig. 4e), reducing the number of DWs in FE. As a result, $C_{DW}$ reaches a maximum and starts decreasing with increasing $V_0$ above $+V_C$ (Fig. 5d).

Similar mechanisms but involving the opposite polarization govern $C_{DW}$ during the backward path. As $V_0$ decreases from 5.5 V to $-5.5$ V, $-P$ domains stabilize, enhancing the response of the $+P$ regions near the DWs to the sinusoidal waveform and undergo domain growth. As $V_0$ decreases further, more DWs participate in this process leading to increase in $C_{DW}$. With further decrease in $V_0$, $-P$ domains nucleate increasing the number of DWs and in turn $C_{DW}$. Eventually, $C_{DW}$ reaches a peak and decreases due to the coalescence of $-P$ domains. These interactions between *P*-switching via domain growth, the change in number of DWs via domain nucleation and coalescence govern the domain wall vicinity response.

## VII. FERROELECTRIC SMALL SIGNAL CAPACITANCE: SUMMARY

Let us summarize the different mechanisms governing the butterfly *C-V* characteristics of the MFIM stacks. The total ferroelectric capacitance ($C_{TOT}$) consists of dielectric ($C_{DE}$) and polarization ($C_P$) capacitance components. $C_P$ component primarily accounts for the butterfly capacitance characteristics of the MFIM, while $C_{DE}$ exhibits an inverted butterfly shape due to in-plane to out-of-plane E-field transformation and vice versa. $C_{DE}$ serves to mildly reduce the range of small-signal capacitance as its trends are opposite to that of $C_P$. $C_P$ is composed of domain bulk ($C_{DB}$) and domain wall ($C_{DW}$) responses, both influenced by the $P$-domain configuration and the voltage-dependent polarization switching mechanisms in the FE layer.

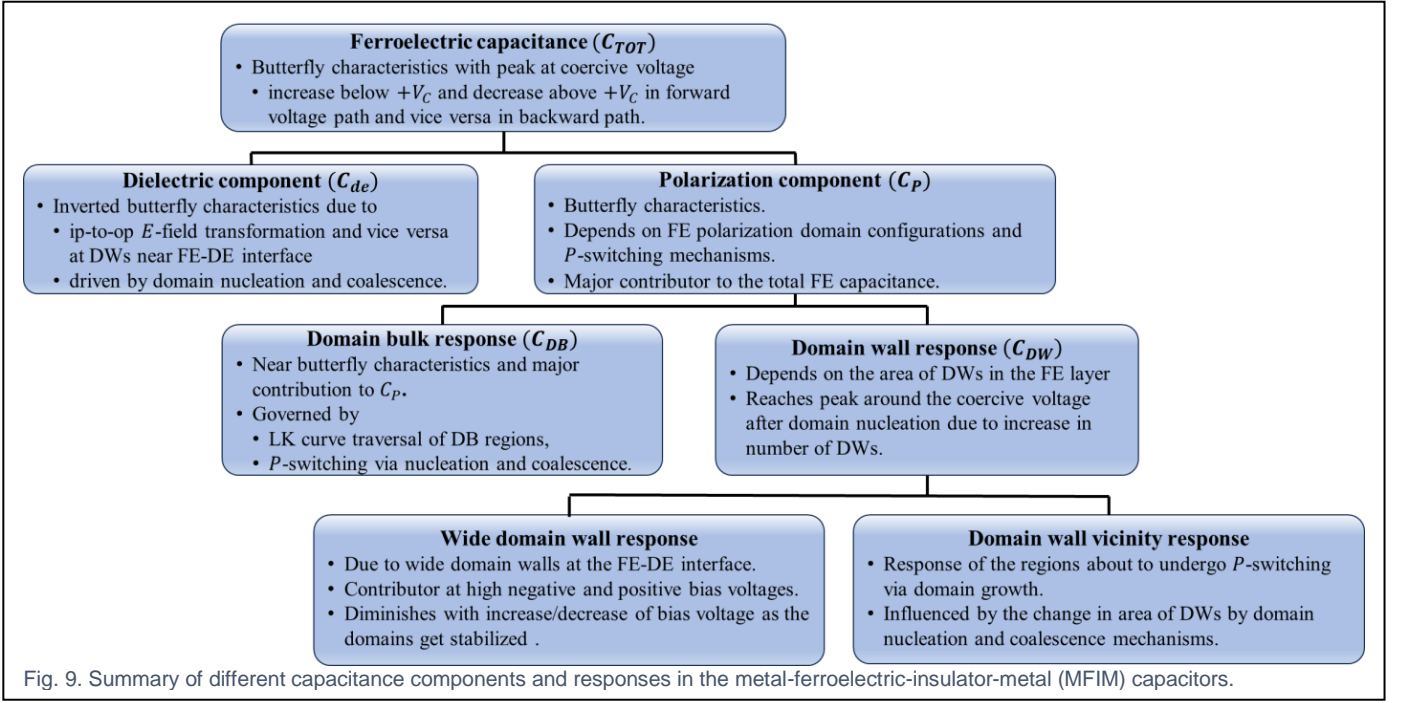

Fig. 9. Summary of different capacitance components and responses in the metal-ferroelectric-insulator-metal (MFIM) capacitors.

Starting at a bias voltage ($V_0$) of $-5.5$ V and moving along the forward voltage path, $C_{TOT}$ increases primarily due to the $C_{DB}$. This increase is attributed to the $P$-$E$ positions of $-P$ DB regions shifting from the low-slope segment of the LK curve towards high slope segment. Additionally, there is a contribution from wide DW response driven by the conical $+P$ domains near the FE-DE interface. However, as the FE layer is negatively polarized with minimal $+P$ domains and DWs, $C_{DB}$ dominates the total capacitance. Further, the wide DW response diminishes as bias voltage increases and the $+P$ domains stabilize into cylindrical shape.

As $+P$ domains stabilize, there is an increase in the contribution from the DW vicinity response, arising from the FE regions near DWs about to undergo domain growth. This response depends significantly on the area of DWs in the FE layer. For the simulated configuration of 10 nm HZO + 2 nm $Al_2O_3$, due to the lower area of DWs, the DW vicinity response is less pronounced than the DB response for $V_0 < +V_C$.

As $V_0$ approaches $+V_C$, the $P$-$E$ positions of $-P$ DB regions reach the high-slope turnaround point of the LK curve, resulting in a peak in the total capacitance near $+V_C$. With further increase in the bias voltage, FE layer undergoes $P$-switching via $+P$ domain nucleation. This process increases the area of DWs in the FE and reduces the domain bulk regions. Consequently, $C_{DB}$ decreases while $C_{DW}$ increases to reach a peak value. At a certain voltage at which $+P$ domain start to coalesce, number of domain walls decrease leading the reducing in $C_{DW}$. At such voltages, $C_{DB}$ is also less as majority of the $+P$ DB regions traverse the low slope regions of the LK curve. This results in the decrease of $C_{TOT}$ above $+V_C$.

Similar mechanisms are observed on the reverse path leading to an initial increase in the capacitance till it reaches a peak, followed by a decrease. To sum up, the voltage dependent interactions between the DB and DW responses, along with the polarization switching via domain growth, nucleation and coalescence gives rise to the butterfly shaped $C$-$V$ characteristics observed in MFIM stacks. An overview of these different capacitance components is summarized in Fig. 9.

## VIII. Scaling Ferroelectric Thickness and Capacitive Memory Window

Let us now examine the impact of scaling the ferroelectric thickness on the small signal $C$-$V$ characteristics. We simulated MFIM stacks with varying HZO thicknesses of 5, 7 and 10 nm and a 2 nm $Al_2O_3$ layer. The bias voltage range was chosen to maintain a constant electric field across the FE layer at the maximum voltage point for all FE thicknesses. It is important to note that we utilize the material and TDGL parameters that are calibrated for 10 nm HZO and 2 nm $Al_2O_3$ MFIM samples for these simulations. However, real world scenarios could introduce changes in the parameter values at different FE thicknesses due to factors such as strain, variations in processing conditions and so on. Therefore, this section focuses on trends in capacitance behavior with scaling of FE thickness rather than absolute capacitance values.

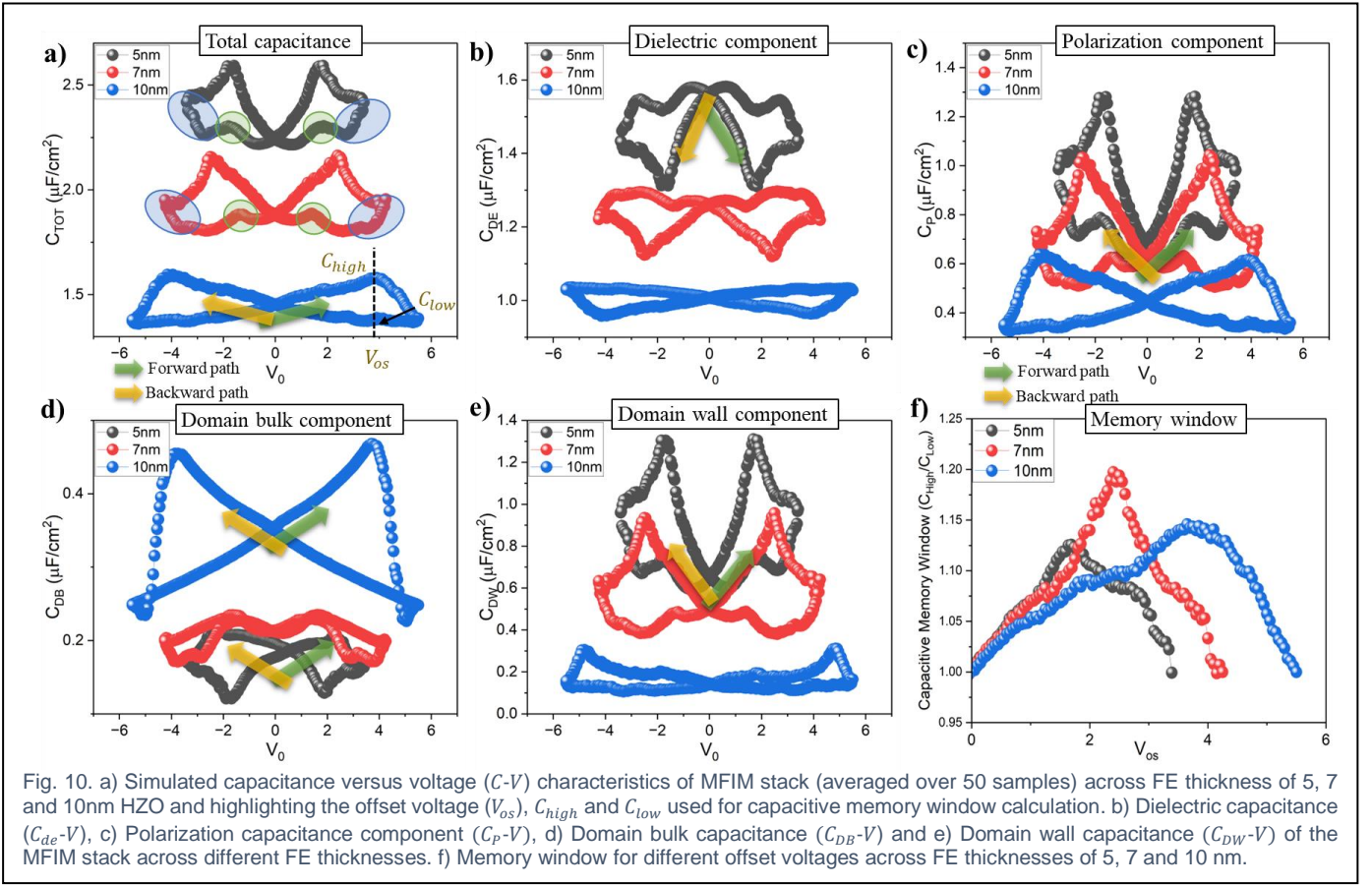

Fig. 10. a) Simulated capacitance versus voltage ($C$-$V$) characteristics of MFIM stack (averaged over 50 samples) across FE thickness of 5, 7 and 10nm HZO and highlighting the offset voltage ($V_{os}$), $C_{high}$ and $C_{low}$ used for capacitive memory window calculation. b) Dielectric capacitance ($C_{de}$-$V$), c) Polarization capacitance component ($C_P$-$V$), d) Domain bulk capacitance ($C_{DB}$-$V$) and e) Domain wall capacitance ($C_{DW}$-$V$) of the MFIM stack across different FE thicknesses. f) Memory window for different offset voltages across FE thicknesses of 5, 7 and 10 nm.

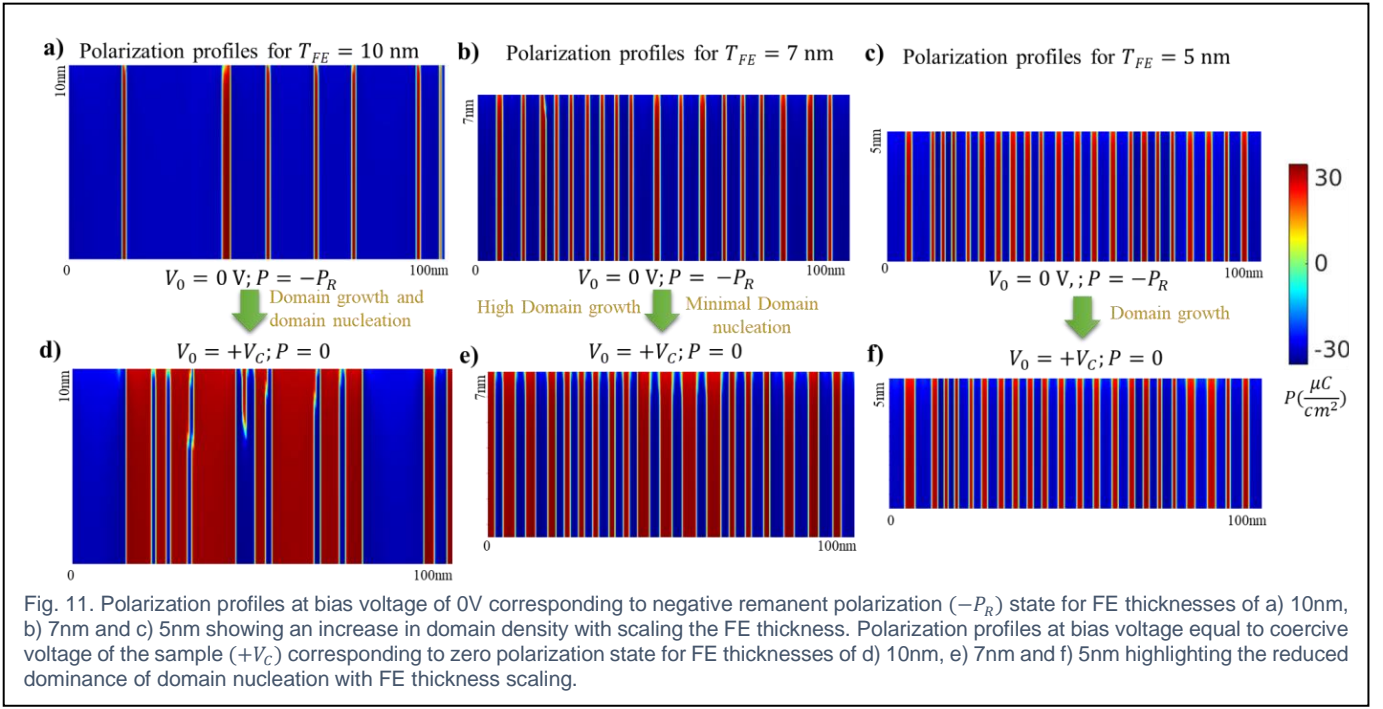

Fig. 11. Polarization profiles at bias voltage of 0V corresponding to negative remanent polarization $(-P_R)$ state for FE thicknesses of a) 10nm, b) 7nm and c) 5nm showing an increase in domain density with scaling the FE thickness. Polarization profiles at bias voltage equal to coercive voltage of the sample $(+V_C)$ corresponding to zero polarization state for FE thicknesses of d) 10nm, e) 7nm and f) 5nm highlighting the reduced dominance of domain nucleation with FE thickness scaling.

The $C$-$V$ characteristics (Fig. 10a) exhibit butterfly shape across different FE thicknesses, showing an increasing trend in the capacitance as FE thickness scales. Dividing the total capacitance, we observe, as before, that the dielectric capacitance $(C_{DE})$ exhibits an inverted butterfly shape, while the polarization capacitance $(C_P)$ shows butterfly characteristics across FE thicknesses (Fig. 10 b, c). Both components exhibit an increasing trend with FE thickness scaling. Additionally, we observe an increase in the range of capacitance across bias voltage in the total capacitance as well as the dielectric and polarization components. These trends in capacitance can be partly attributed to the reduction in total device thickness. However, our analysis reveals an additional and a significant contribution from the changes in the underlying polarization domain configuration with FE thickness scaling.

Scaling FE thickness increases the polarization domain density in the FE layer (Fig. 11 a, b, c) due to interplay between gradient and electrostatic energy, as demonstrated by previous studies [38, 46]. The increased domain density increases the area of DWs in the FE layer and reduces the DB regions. Consequently, we observe a decrease in $C_{DB}$ with FE thickness scaling (Fig. 10d) and an increase in $C_{DW}$ (Fig. 10e). The increase in $C_{DW}$ is higher than decrease in $C_{DB}$ because the domain wall regions fall into the high-slope segment of the LK curve, whereas the DB regions typically fall into low-slope segment of the LK curve, except on the verge of domain nucleation. As a result, $C_{DW}$ contributes significantly to $C_P$ at lower FE thicknesses due to which we observe an increase in $C_P$ with FE thickness scaling.

Further, the increase in domain density with the reduction in FE thickness alters the $P$-switching mechanism from a combination of domain growth and domain nucleation at higher FE thicknesses to domain growth dominated at lower sFE thicknesses (Fig. 11) [46]. As a result, for low FE thickness, there is an increased number of domain walls on the verge of undergoing domain growth (at a bias voltage associated with $P$-switching). Therefore, we observe an increase in the maximum $C_{DW}$ (around coercive voltage of the sample) and in turn an increase in the range of $C_{DW}$ and $C_P$.

Next, let us discuss the trends of $C_{DE}$ across FE thickness. The increased domain density with FE thickness scaling leads to an increase in the in-plane $E$-field (i.e. reduction in out-of-place E-fields), which tends to reduce the charge response of the ferroelectric. As a result, $C_{DE}$ shows a lower sensitivity to FE thickness compared to the scenario when the polarization domain effects are absent i.e., when the capacitance change is strictly due to geometry change, as in a linear capacitance.

Additionally, we observe increased range of $C_{DE}$ across bias voltages at lower FE thicknesses (7 and 5 nm), which can be understood as follows. Let us focus on the forward path and start at the most negative bias voltage. Even with the increased domain density due to FE thickness scaling, majority of the $E$-field is in out-of-plane direction due to the asymmetry between the magnitudes of $+P$ and $-P$ domains. As $V_0$ increases, this asymmetry reduces, resulting in the transformation of the $E$-field from out-of-plane to in-plane direction. Due to the higher domain density and in turn higher area of DWs, we see an increased op-to-ip transformation compared to 10 nm. As a result, $C_{DE}$ reduces more for lower FE thicknesses, leading to higher range of $C_{DE}$ with FE thickness scaling.

However, we observe certain deviations in the $C$-$V$ characteristics from the traditional butterfly curves at FE thicknesses of 7 and 5 nm (highlighted in Fig. 10a). Firstly, there is an initial reduction in capacitance (highlighted in blue in Fig. 10a) with an increase (decrease) in $V_0$ from the most negative (positive) bias voltage during the forward (backward) voltage paths. This reduction is primarily attributed to the wide domain wall response at the FE-DE interface. Recall the formation of conical $+P$ domains at highly negative $V_0$ and that the $P$-$E$ position of $+P$ domain point at the FE-DE interface falls along the high-slope segment of the LK curve (refer

Section VI.B.i). As $V_0$ increases, this point shifts toward the low-slope segment, reducing its response to the sinusoidal waveform. Due to the large number of DWs at lower FE thicknesses, this effect is more pronounced and leads to a reduction in $C_{TOT}$.

Secondly, we observe small bumps in the capacitance (highlighted in green in Fig. 10a) during the transition of $V_0$ from most negative (positive) voltage to 0V. This can also be attributed to the wide-DW response at the FE-DE interface. At high negative $V_0$, the "softer" DW regions associated with the conical $+P$ domains have their $P$-$E$ position along the moderate-slope segment of the LK curve (refer Section VI.B.i). As $V_0$ increases, this $P$-$E$ position transitions along the moderate-slope segment and reaches the high-slope segment of the LK curve, increasing its response to the sinusoidal waveform. Due to higher number of DWs at lower FE thicknesses, this results in the increase of $C_{TOT}$ leading to a small bump. With further increase in $V_0$, the $P$-$E$ position transitions from the high-slope segment to low-slope segment leading to decrease of $C_{TOT}$ in the bump. This phenomenon needs further investigation considering more rigorous 3D simulations and experimental characterizations at scaled FE thickness.

We further explore the impact of increased domain density and FE thickness scaling on the capacitive memory window, a crucial metric for capacitive compute-in-memory (CiM) applications [17], [18], [19]. Ferroelectric capacitors are utilized in capacitive CiM applications by creating a difference between the capacitances of the forward and backward paths (or the low and high capacitance states) at zero bias voltage. This difference is achieved either by introducing a non-zero offset voltage ($V_{os}$) across the FE capacitor or by incorporating traps, which cause asymmetry in the butterfly $C$-$V$ characteristics [18]. For our analysis, we consider the former method.

We vary the offset voltage (Fig. 10a) and examine the trends in the capacitive memory window (MW), defined as the ratio between the high capacitance state ($C_{high}$) and the low capacitance state ($C_{low}$), as marked in Fig. 10a [17]. We observe a non-monotonic trend in the capacitive MW versus the offset voltage for different FE thicknesses. Increasing $V_{os}$ from 0 V, the capacitive MW initially increases with the offset voltage ($V_{os}$) and then decreases (Fig. 10f) leading to a peak in the MW at certain $V_{os}$. This characteristic of the MW is primarily due to the butterfly shape of the FE $C$-$V$ characteristics. Further, we observe a non-monotonic trend in the peak capacitive MW with the scaling of FE thickness (Fig. 10f). There is an increase in the peak MW as FE thickness scales from 10 to 7 nm attributed to the increase in the range of $C_{TOT}$. This is followed by a decrease in peak MW as FE scales from 7 to 5 nm. Although the range of $C_{TOT}$ increases at 5nm, this decrease is observed due to the bump in the butterfly curves, discussed in the previous paragraph. The peak of $C_{high}$ occurs at $V_{OS}$ which aligns with the bump in the $C_{low}$ branch of the butterfly curve. This reduces the peak $C_{high}/C_{low}$. As mentioned before, the bump in the butterfly curves and this reduction in MW at 5nm needs further investigation.

Nevertheless, the dependence of capacitive MW on FE thickness presented here highlights the important role of $P$-domain configurations in the capacitance behavior of the MFIM stacks. If the capacitance changes were only due to geometry (thickness) scaling, the MW would remain constant across different FE thicknesses.

## IX. CONCLUSION

In summary, we have investigated the butterfly ferroelectric capacitance-voltage characteristics. We have analyzed the dielectric and polarization capacitance components and the physical mechanisms governing them. Our analysis revealed three key mechanisms responsible for the polarization capacitance component. The domain bulk region is dominant in the regions away from the domain walls and is the major contributor to the ferroelectric capacitance at higher FE thicknesses. The wide domain wall response at the FE-DE interface is prevalent only at highly negative and positive bias voltages. The domain wall vicinity response is dominant near the domain wall regions of the FE layer. The contribution of the domain wall response becomes more important at scaled FE thicknesses. We also discuss the impact of FE thickness scaling and increased domain density on the capacitance-voltage characteristics and in turn on the capacitive memory window, showing a non-monotonic trend in the peak capacitive memory window as a function of FE thickness.